\documentstyle[twocolumn,psfig,prb,aps,eqsecnum]{revtex}

\begin{document}
\draft
\preprint{scheduled to appear in Phys. Rev. B, Feb. 15 (1998)}
\title{
Structure and dynamics of Rh surfaces
}
\author{Jianjun Xie and  Matthias Scheffler}
\address{
Fritz-Haber-Institut der Max-Planck-Gesellschaft,\\
Faradayweg 4-6, D-14195 Berlin-Dahlem, Germany
}
\date{August 7, 1997}
\maketitle
\begin{abstract}
Lattice relaxations, surface phonon spectra, surface energies, and
work functions are calculated for Rh(100) and Rh(110) surfaces
using density-functional theory and the full-potential
linearized augmented plane wave  method.
Both, the local-density approximation
and the generalized gradient approximation to the exchange-correlation 
functional are considered.
The force constants are obtained from the directly calculated
atomic forces, and the temperature dependence of the surface relaxation
is evaluated by  minimizing the free energy of the system.
The anharmonicity of the atomic vibrations is taken into account within
the quasiharmonic approximation.
The importance of contributions from different phonons 
to the surface relaxation is analyzed.
\end{abstract}


\section{Introduction}
\label{sec:level1}

Experimentally, the structure of a solid surface can be determined by
LEED~\cite{vanHove86}, ion scattering~\cite{Varlas86},
X-ray diffraction~\cite{Rieder86}, and 
helium-atom scattering~\cite{Rieder86}. Surface phonons 
can be measured by high-resolution electron energy-loss spectroscopy 
(EELS)~\cite{Ibach82} and inelastic helium-atom
scattering~\cite{Toennies}. 
It is well understood that the first interlayer
separation of metal surfaces is typically contracted at room 
temperature~\cite{Schef92}, and only few exceptions to this ``rule''
have been discovered so far. Due to the 
anharmonicity of the inter-atomic potentials, both the surface interlayer spacing 
and the bulk lattice constant depend on temperature. Furthermore,
because of lower symmetry, the anharmonicity at the surface is
expected to be more significant than that in the bulk.

First-principles calculations of lattice
relaxations~\cite{Schef92,Rodach,Feibelman,Freeman,Kleinman,Eichler96}
and
surface phonons for metals~\cite{Ho86,Ho93,Bohnen96,Eguiluz} typically employ
the density-functional theory (DFT)~\cite{Review}. In most previous work,
the equilibrium geometry was determined by minimizing  the
total energy of the system (we will call this the ``static equilibrium"). 
We note that the surface structure obtained in this way corresponds
to $T = 0$ K and neglects the effects of zero-point vibrations.
With respect to the latter we note that for example for the crystal
bulk zero-point vibrations give rise to an
increase of the lattice constant by about 0.2-0.5\%, compared to the 
geometry determined by the total-energy minimum~\cite{Williams,Biernacki89}.
Starting from the ``static equilibrium'' geometry, surface phonons 
were calculated either directly (using the supercell 
approach)~\cite{Ho86,Ho93,Bohnen96} or by 
applying density-functional perturbation theory~\cite{Eguiluz,Baroni}.

At finite temperature, the equilibrium geometry is 
determined by the minimum of the free energy, which, in addition to 
the total energy, includes the contribution from lattice vibrations.
Typically this gives rise to a lattice expansion, but it is also
possible that contractions occur \cite{Biernacki89}. Because 
anharmonicity is more pronounced at the surface than in the bulk 
an increased interlayer spacing of the top layer, $d_{12}$,
is to be expected~\cite{Cao,Gust}.
Recently, Narasimhan and 
Scheffler~\cite{Nara} and Cho and Scheffler~\cite{Cho}
performed a theoretical study of the thermal expansion of Ag(111) 
and Rh(100).
The free energy was calculated by including the contribution of 
phonons in the quasiharmonic approximation~\cite{Biernacki89,Wette69}. 
The temperature dependence of $d_{12}$ was obtained by minimizing the free
energy of the considered slab with respect to $d_{12}$. In these 
studies~\cite{Nara,Cho}, 
as a first approximation, only the top layer was allowed  
to vibrate as a whole parallel and perpendicular to the surface.
The effective vibrational ``modes'' considered in
these works correspond to modes at the Brillouin zone center for 
the top layer vibrating on a rigid substrate.
No information about the true surface phonon spectrum was obtained,
and thus, not even an attempt was made to discuss
how a good-quality summation over the phonon Brillouin zone
may affect the result. In fact a better treatment is quite
elaborate as it requires the diagonalization of the dynamical matrix of
the whole slab.
In order to improve on the previous calculations~\cite{Cho},
we present in this paper a theoretical study of 
the structure and dynamics of Rh(100) and Rh(110) surfaces by
including vibrations of the whole slab. 
The interplanar force constants are obtained 
from the directly calculated atomic forces using the
full-potential linearized augmented plane wave (FP-LAPW) method~\cite{Wien95,Force}.
It has been shown~\cite{Ho86,Ho93} that the surface phonon modes 
for wave vectors at high-symmetry points of the surface
Brillouin zone (SBZ) can be determined from the knowledge of
interplanar force constants. 
In the present work, these force constants and the corresponding 
 surface phonons are calculated as a function
of the interlayer spacing at the surface. 
The temperature dependence of  surface relaxation is then determined by minimizing
the free energy 
within the quasiharmonic approximation. 
Contributions from different phonon modes to the vibrational
free energy will be  examined. 
 In Section~\ref{sec:level2}, we will describe
some details of the calculation method. The results are presented
in Section~\ref{sec:level3}.
 
\section{Method of calculation}
\label{sec:level2}
\subsection{General theory}

We consider a crystal slab consisting of
a finite number of atomic layers perpendicular to the $z$
direction, and of infinite extension in the $x$ and $y$ directions.  
In the quasiharmonic approximation, the free-energy function of the considered
system can be written as~\cite{Biernacki89} 
\begin{eqnarray}
F(\{d_{ij}\},T)&=&\Phi(\{d_{ij}\})+F_{\rm vib}(\{d_{ij}\},T) \nonumber\\
&=&\Phi(\{d_{ij}\})+k_{\rm B}T\sum_{{\bf q}_\parallel}\sum_{p=1}^{3N} \nonumber\\
&{\rm ln}&\left\{2\sinh\left(\frac{\hbar\omega_{p}({\bf q}_\parallel,\{d_{ij}\})}
{2k_{\rm B}T}\right)\right\} \quad,
\end{eqnarray}
where $\Phi$ is the static total energy which can  
be obtained  by first-principles calculations, 
$k_{\rm B}$ and $\hbar$ are the Boltzmann and the Planck constants,
and \{$d_{ij}$\} is the set of inter-layer distances, $d_{12}$, $d_{23}$, $\cdots$,
between layers 1 and 2, 2 and 3, etc.
The vibrational free energy is denoted as $F_{\rm vib}$,
and $\omega_{p}({\bf q}_\parallel,\{d_{ij}\})$ is the frequency of the 
$p$-th mode for a given wave vector {\bf q}$_\parallel$, evaluated at the geometry
defined by \{$d_{ij}$\}; and $N$ is the number of atoms in the slab.
Anharmonicity of the inter-atomic potentials is included in this description
because $\Phi$ and the vibrational frequencies $\omega_p$ depend
on the inter-layer distances \{$d_{ij}$\}.
The free energy and the equilibrium distances \{$d_{ij}^{0}$\} at a given temperature
are determined by the minimum of $F(\{d_{ij}\},T)$ with respect to \{$d_{ij}$\}. 

In the present work, we have calculated  the vibrational frequencies and the 
corresponding free energy
using the slab model~\cite{Wette71} and density-functional theory. 
We expand the static total energy, $\Phi$,
of the system in a Taylor series
\begin{eqnarray}
\Phi & = & \Phi_{0} + \sum_{{\bf l},\alpha}\phi_{\alpha}({\bf l})
u_{\alpha}({\bf l})\nonumber\\
&+&\frac{1}{2}\sum_{{\bf l},\alpha}
\sum_{{\bf l}^\prime, \beta}\phi_{\alpha\beta}({\bf l},{\bf l}^{\prime})
u_{\alpha}({\bf l})u_{\beta}({\bf l}^{\prime})+...\quad,
\end{eqnarray}
where $u_{\alpha}({\bf l})$ is the $\alpha$ component ($\alpha=x, y, z)$
of the displacement of the {\bf l}-th atom from its  mean position
${\bf R}_{0}({\bf l})$ (taking the thermal expansion of the lattice into account). 
The instantaneous position of an atom is
given by {\bf r}({\bf l})={\bf R}$_{0}$({\bf l})+{\bf u}({\bf l}). 
We consider here the monatomic lattice, and the set of integers
{\bf l}=$(l_{1}, l_{2}, l_{3})$, which  specify a particular
atom, has the following meaning: $l_{3}$ labels the crystal
planes lying parallel to the surface, and $l_{1}$, $l_{2}$ specify the points in
the two-dimensional lattice which spans a plane.
In the quasiharmonic approximation the equations of motion are
\begin{equation}
 M\frac{d^{2}}{dt^{2}}u_{\alpha}({\bf l})=-\sum_{{\bf l}^{\prime},
\beta}\phi_{\alpha\beta}({\bf l},{\bf l}^{\prime})u_{\beta}
({\bf l}^{\prime})\quad ,
\end{equation}
where $M$ is the atomic mass, $\phi_{\alpha\beta}({\bf l},{\bf l}^{\prime})$ 
is the force constant which is defined by
\begin{equation}
\phi_{\alpha\beta}({\bf l,l}^{\prime})=\left(\frac{\partial^{2}\Phi}
{\partial u_{\alpha}
({\bf l})\partial u_{\beta}({\bf l}^{\prime})}\right)_{0}\quad .
\end{equation}
The subscript ``0" in Eq.(2.4) indicates that the force constants 
$\phi_{\alpha\beta}({\bf l,l}^{\prime})$ are to be evaluated at the
mean positions of the atoms, rather than at the positions of the
static equilibrium. 
Due to the two-dimensional translational property of the slab, 
the normal mode solutions to Eq.(2.3) have the form~\cite{Wette71}
\begin{eqnarray}
u_{\alpha}({\bf l})&=&M^{-1/2}Q_{0}\zeta_{\alpha}(l_{3})
{\rm exp}\{i[{\bf q}_\parallel\cdot {\bf R}_{0||}({\bf l}_\parallel) \nonumber\\
&+&{\bf q}_\parallel\cdot 
{\bf R}_{0||}(l_{3})-\omega t]\} \quad ,
\end{eqnarray}
where $Q_{0}$ is the vibrational  amplitude,
$\zeta_{\alpha}(l_{3})$ is a polarization vector which will turn out to be
the eigenvectors of the dynamical matrix, and $\omega$ is
 the
vibrational frequency. Following the notation of Ref.~\cite{Wette71}
, we have {\bf q}$_\parallel=(q_{x}, q_{y}), {\bf R}_{0||}=(x,y), 
{\bf R}_{0||}({\bf l})={\bf R}_{0||}({\bf l}_\parallel)+{\bf R}_{0||}(l_{3}),
{\bf l}_\parallel=(l_{1},l_{2}).$ 
Insertion of Eq. (2.5) into Eq. (2.3) leads to the eigenvalue equation
\begin{equation}
\sum_{l_{3}^{\prime},\beta}D_{\alpha\beta}(l_{3},l^{\prime}_{3};
{\bf q}_\parallel)\zeta_{\beta}(l_{3}^{\prime};{\bf q}_\parallel)=
\omega^{2}({\bf q}_\parallel)
\zeta_{\beta}(l_{3};{\bf q}_\parallel) \quad ,
\end{equation}
where the elements of the dynamical matrix  are defined by
\begin{eqnarray}
D_{\alpha\beta}(l_{3},l^{\prime}_{3};{\bf q}_\parallel)&=&\frac{1}{M}
\sum_{{\bf l}_\parallel^\prime}\phi_{\alpha\beta}(l_{3},l_{3}^{\prime}
,{\bf l}_\parallel^{\prime}-{\bf l}_\parallel)\nonumber\\
& &\times {\rm exp}\{i{\bf q}_\parallel\cdot [{\bf R}_{0||}
({\bf l}^{\prime}_\parallel-{\bf l}_\parallel)\nonumber\\
&+&{\bf R}_{0||}
(l_{3}^{\prime})-{\bf R}_{0||}(l_{3})]\} \quad .
\end{eqnarray}
Once the force constants are obtained, the frequencies and polarization
vectors for all modes in the slab for a given  ${\bf q}_\parallel$ can 
be obtained by diagonalization of a $3N\times 3N$ matrix.
 
\subsection{Density functional theory calculation}
\label{sec:level2b}

In the present paper, 
the considered systems Rh(100) and Rh(110) are modeled
by a periodic slab consisting of 7 layers of Rh and a vacuum
region with the same thickness.  We employ
density-functional theory and the
FP-LAPW method~\cite{Wien95,Force}.
The exchange-correlation functional is treated by the 
local-density approximation (LDA)~\cite{LDA92} as well as the generalized gradient
approximation (GGA)~\cite{GGA92}.
We first calculate the lattice relaxations, surface energies and work functions
of Rh surfaces at $T=0$ K neglecting the influence of
zero-point vibrations. 
The geometry is optimized by a damped
molecular dynamics allowing the top two layers on both sides of slab 
to relax. The remaining atoms are kept at the bulk lattice sites.
The energy cut-off for the FP-LAPW basis is taken to be
15 Ry and for the wave functions inside the muffin-tin spheres
angular momenta are taken into account up to 
$l_{max}^{wf}=8 $. 
The cut-off energy for the potential is taken as $G^{2}= 100$ Ry.
For the ${\bf k}_\parallel$-sampling, we use a uniform mesh of 15 points in the 
irreducible part
of SBZ of the ($1\times 1$) surface displaced from $\overline{\Gamma}$.

 In order to calculate phonon frequencies of the slab, an extension of the
``frozen phonon" method~\cite{Ho86} is used. 
For a selected ${\bf q}_\parallel$ in the surface Brillouin
zone, we can calculate all the interplanar force constant matrices by distorting
the slab in an appropriate way. The obtained force constants allow us to set
up the dynamical matrix and solve for the eigenvalues and eigenvectors of the
system. The interplanar force constants coupling planes $l_{3}$ and 
$l_{3}^{\prime}$ are given by
\begin{eqnarray}
\phi^{p}_{\alpha\beta}(l_{3},l_{3}^{\prime}; {\bf q}_\parallel)& =&
\sum_{{\bf l}^{\prime}_\parallel}\phi
_{\alpha\beta}(l_{3},l_{3}^{\prime},{\bf l}^{\prime}_\parallel-{\bf
l}_\parallel) \nonumber \\
&\exp&[i{\bf q}_\parallel\cdot
{\bf R}_{0||}({\bf l}^{\prime}_\parallel-{\bf l}_\parallel)]\quad .
\end{eqnarray}
In the harmonic or quasiharmonic approximation, the interplanar force constants,
$\phi^{p}_{\alpha\beta}(l_{3},l_{3}^{\prime}; {\bf q}_\parallel)$,
are related to the forces by
\begin{equation}
\phi_{\alpha\beta}^{p}(l_{3},l_{3}^{\prime};{\bf q}_\parallel)=
-\frac{\partial F_{\alpha}(l_{3},
{\bf q}_\parallel)}{\partial u_{\beta}({\bf l}_\parallel,l_{3}^{\prime})}\approx 
-\frac{\Delta F_{\alpha}(l_{3},{\bf q}_\parallel)}
{u_{\beta}({\bf l}_\parallel,l_{3}^{\prime})} \quad ,
\end{equation}
where $\Delta F_{\alpha}(l_{3},{\bf q_\parallel})$ is the $\alpha$ component of
the force difference at atomic layer $l_{3}$ under a distortion
of layer $l_{3}^{\prime}$ in accordance with ${\bf q}_\parallel$, 
$u_{\beta}({\bf l}_\parallel,l_{3}^{\prime})$ is the $\beta$ component of
 atomic displacement at 
$({\bf l}_\parallel,l_{3}^{\prime})$. The displacement for other atoms 
in layer $l_{3}^{\prime}$ is given by
\begin{equation}
{\bf u}({\bf l}_\parallel^{\prime},l_{3}^{\prime})=
{\bf u}({\bf l}_\parallel,l_{3}^{\prime})
\exp[i{\bf q}_\parallel \cdot {\bf R}_{0||}
({\bf l}_\parallel^{\prime}-{\bf l}_\parallel)]\quad .
\end{equation} 
The dynamical matrix of Eq.(2.7) can then be written as
\begin{eqnarray}
D_{\alpha\beta}(l_{3},l_{3}^{\prime};{\bf q}_\parallel)&=&
\frac{1}{M}\phi_{\alpha\beta}^{p}
(l_{3},l_{3}^{\prime};{\bf q}_\parallel) \nonumber\\
&\exp&\{i{\bf q}_\parallel\cdot
 [{\bf R}_{0||}(l_{3}^\prime)
-{\bf R}_{0||}(l_{3})]\} \quad .
\end{eqnarray}

In order to calculate the thermal expansion, we need to know
the interplanar force constants as a function of the interlayer distance. 
In most cases, the thermal expansion of the lattice at the surface
along $z$ direction is expected to be more significant than the expansion
of the bulk lattice constant. The latter also defines the expansion of 
interatomic distance in the $x$-$y$ plane. We therefore calculate the 
surface interplanar force constants as a function of $d_{12}$. 
Furthermore, we calculate the interplanar force constants for the crystal
bulk. From all the interplanar force constants, the dynamical matrix
(as a function of $d_{12}$) of a thick slab can be constructed.
In fact, for the phonon calculations it is important
to use a significantly thicker slab than for the electronic structure
calculations, in order to obtain a reliable description of surface
modes, surface resonances and the surface phonon density of states
\cite{Ho86,Wette71}.
We therefore calculate the surface phonon spectra for Rh(100) and
Rh(110) for a 201-layer thick slab.
The surface relaxation $d_{12}(T)$ is then obtained 
by minimizing the free energy of this slab for various temperatures.

\section{ Results}
\label{sec:level3}
For Rh bulk the theoretically obtained lattice constant 
(neglecting the influence of zero-point vibrations) is
$a^{\rm th}=3.79$ \AA\, for DFT-LDA and $a^{\rm th}=3.83$ \AA\,
for DFT-GGA.
The FP-LAPW parameters employed for these calculations
are the same as those for the surface calculations
listed in Section~\ref{sec:level2b}, except that for the bulk study
we use a mesh of 72 {\bf k}-points in the irreducible 
part of the fcc Brillouin zone.
The experimental value is $a^{\rm exp}=3.80 {\rm \AA}$~\cite{Kittel86}.
The surface properties of Rh are calculated with the theoretical
bulk lattice constant. In Table \ref{table1},
we present the results of surface relaxations, work functions and
surface energies of Rh(100) and Rh(110) calculated
in the  LDA and GGA neglecting the zero-point vibration.
Results from other calculations and the experiments
are also listed for comparison. The present
results are in good agreement with those of the previous 
calculations~\cite{Schef92,Eichler96,Cho,Stokbro}. 
In the following calculations of the surface dynamics, we will use the GGA. 
The surface interplanar force constant will be calculated as
a function of $d_{12}$, while the interlayer spacing $d_{23}$
is kept as the static equilibrium value.

\begin{figure}
\psfig{figure=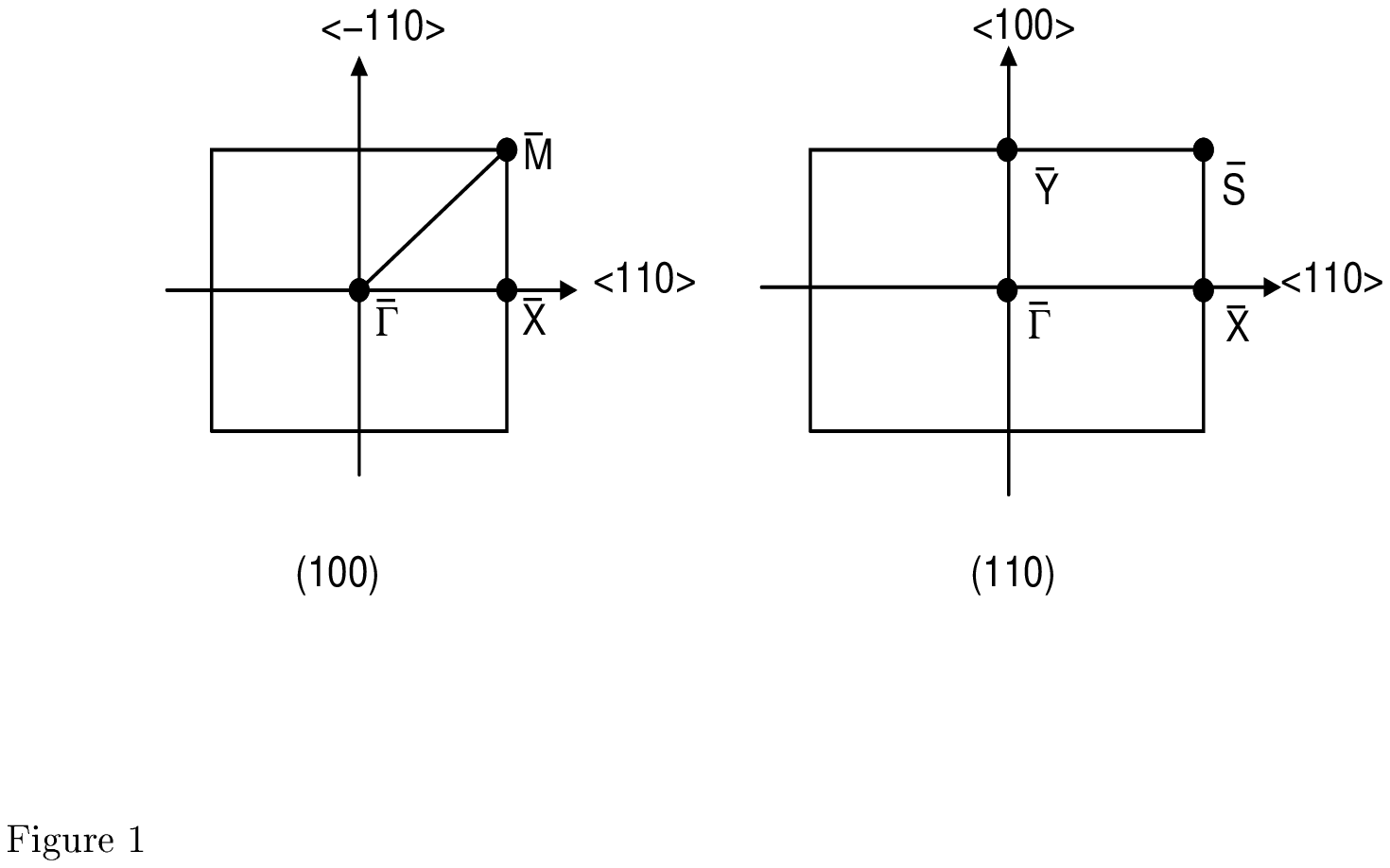,width=8cm}
\caption{Surface Brillouin zones for fcc (100) and (110) surfaces.
High symmetry points are indicated.}
\label{figure1}
\end{figure}

The calculated interplanar force constants coupling the surface layer
with other layers in Rh(100) slab are given in Table~\ref{table2}.
Three high symmetry surface wave vectors are considered
(see Fig.~\ref{figure1}): $\overline{\Gamma}$, 
$\overline{\rm X}$ and $\overline{\rm M}$. 
These results are obtained by distorting the slab around the 
zero-force geometry at $T=0$ K, which is given in Table \ref{table1}.
The corresponding results for the interior of the slab are 
given as well. 
The interplanar force constant matrices for the Rh(110) surface
are given in Table~\ref{table3}. Four wave vectors at $\overline{\Gamma}$,
$\overline{\rm X}$, $\overline{\rm Y}$, and $\overline{\rm S}$ points
in the SBZ are considered (see Fig.~\ref{figure1}). 
It is not surprising to see in Table~\ref{table2} and Table~\ref{table3} that
the intralayer part of the force constants  
($\phi_{\alpha\beta}^{p}(1,1)$) are most significantly changed at the surface
compared to that in the bulk  because
of the truncation of surface. At the same time, the interlayer part of
force constants at surface ($\phi_{\alpha\beta}^{p}(1,l_{3}), l_{3}=2, 3$) 
are also modified due to the surface relaxations.  For most of the matrix
elements in Table~\ref{table3}, there is a trend for an enhancement in
magnitude for force constants coupling the surface layer to its neighbors
compared with the bulk force-constant matrix elements. This is in agreement
with the findings of Al(110)~\cite{Ho86}.  

\begin{figure}
\psfig{figure=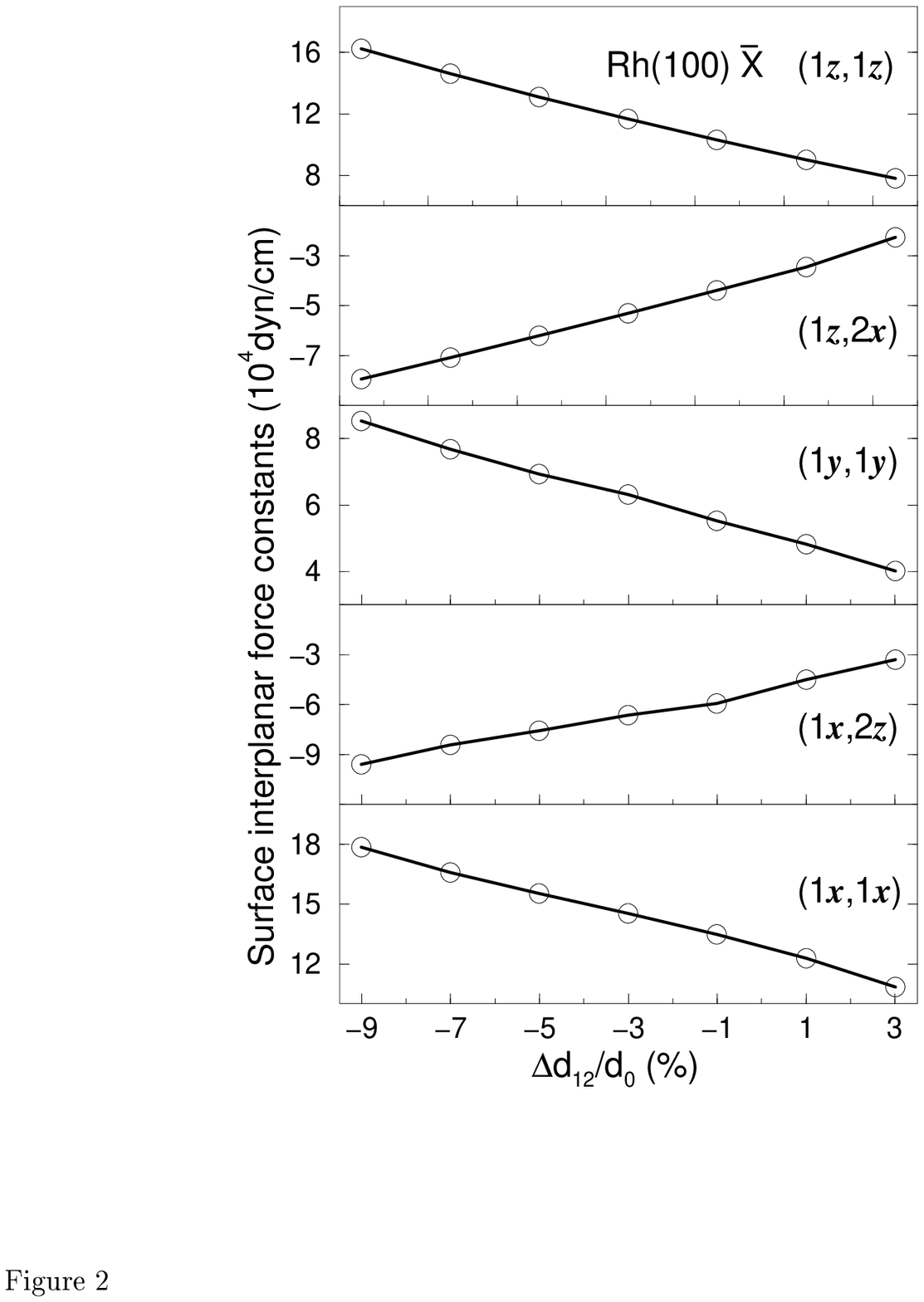,width=8cm}
\caption{Variation of the surface interplanar force constants
at $\overline{\rm X}$ with $\Delta d_{12}/d_{0}$ for Rh(100).}
\label{figure2}
\end{figure}

Anharmonicity is included in our studies by calculating
the force constants as a function of the interlayer distance $d_{12}$. 
Figure \ref{figure2}
shows the variation of the surface interplanar  force constants at the
$\overline{\rm X}$ point for Rh(100). It can be clearly seen that 
the interlayer force constants decrease monotonically as $d_{12}$ changes 
from $ 9 \%$ contraction to $ 3 \%$ expansion. This ``softening'' of the
force constants reflects the anharmonicity of the bond strength at
this surface. We also note that the intralayer
force constants of the second layer ($\phi_{\alpha\beta}^{p}(2,2)$ )
are also significantly softened at the
$\overline{\rm X}$ and $\overline{\rm M}$ points for Rh(100), and
at the $\overline{\rm X}$, $\overline{\rm Y}$,
and $\overline{\rm S}$ points for Rh(110), with the increase of $d_{12}$.
While at $\overline {\Gamma}$,
there are no intraplanar vibrations,
the ``softening'' of $\phi_{\alpha\beta}^{p}(2,2)$ is already included in
the surface interplanar force constant $\phi_{\alpha\beta}^{p}(1,2)$.
In the present work, the changes of the intraplanar force constants for
the second layer with  $d_{12}$ are taken into account.

\begin{figure}
\psfig{figure=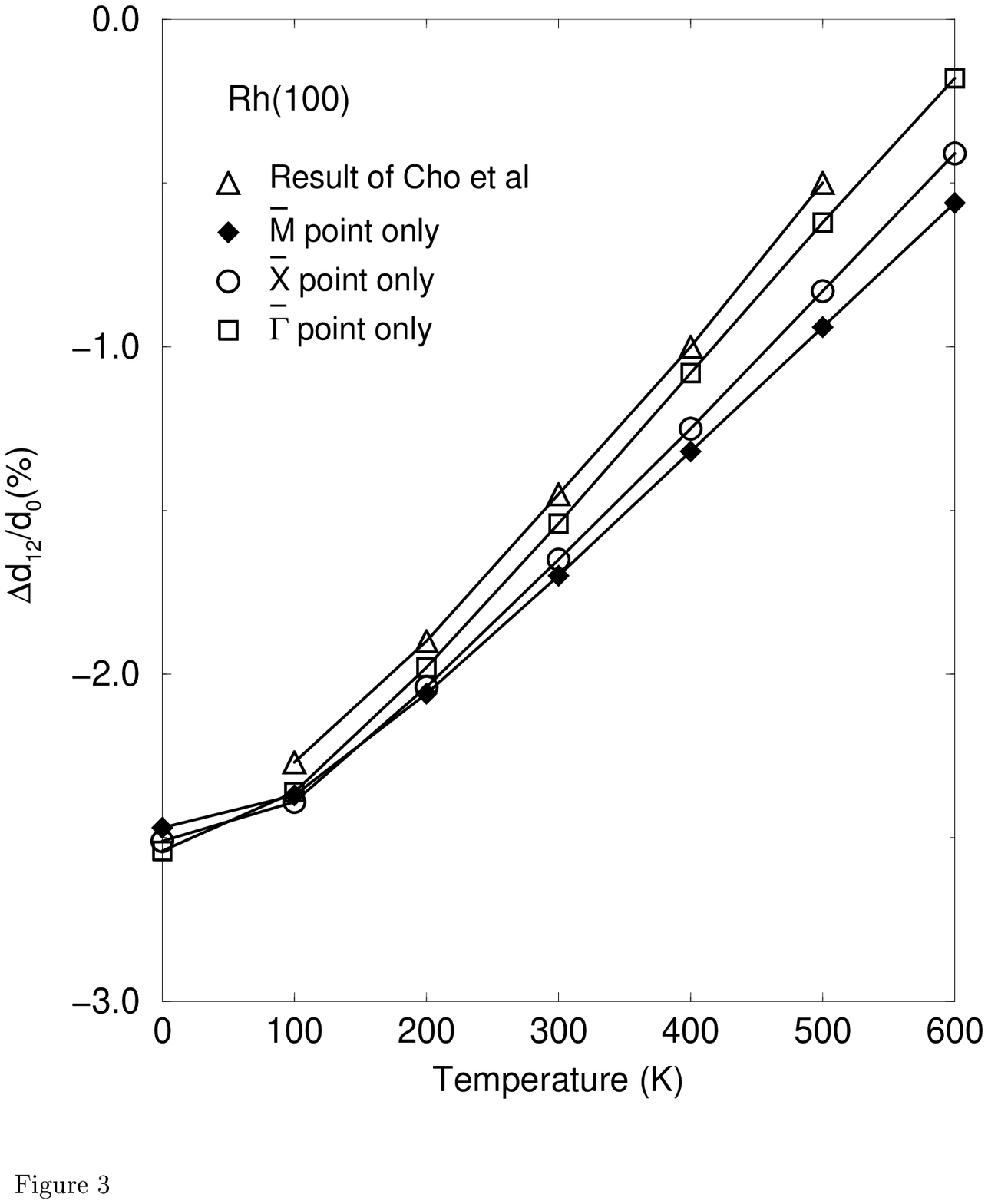,width=8cm}
\caption{Top-layer relaxation of Rh(100) as a function of temperature
calculated in different approximations, i.e., 
including different vibration modes. 
The results of Cho
and Scheffler~{\protect \cite{Cho}} are  shown for comparison.}
\label{figure3}
\end{figure}

Having obtained the force constants for different $d_{12}$, we calculate
the phonon frequencies of the slab for each $d_{12}$. 
Then, using Eq. (2.1), the temperature
dependent top-layer relaxations for Rh(100) and Rh(110) are determined
by minimizing the free energy with respect to $d_{12}$.
Figure~\ref{figure3} shows the variation of $\Delta d_{12}/d_{0}$
 for Rh(100) with temperature
in the case in which  vibrations are calculated from different wave 
vectors ${\bf q}_\parallel$ in the SBZ.
The results of Cho and Scheffler~\cite{Cho} are also given for
comparison. 
It can be seen that when only the $\overline{\Gamma}$ point
vibration is taken into account, the variation of  $\Delta d_{12}/d_{0}$
with temperature is in close agreement with the results of Ref.~\cite{Cho}.
The small difference is due to the fact that only the top layer
was allowed to vibrate in Ref.~\cite{Cho}, while in the present work,
vibrations of the whole 201-layers thick slab are included.  
At low temperature (below 300 K), the vibrations from
the $\overline{\Gamma}$, $\overline{\rm X}$, and $\overline{\rm M}$
points contribute with similar importance to the surface expansion.
The differences between the contributions of the different
${\bf q}_\parallel$ vectors becomes significant at high temperature
because of the different frequency distributions between $\overline{\Gamma}$,
$\overline{\rm X}$ and $\overline{\rm M}$.

\begin{figure}
\psfig{figure=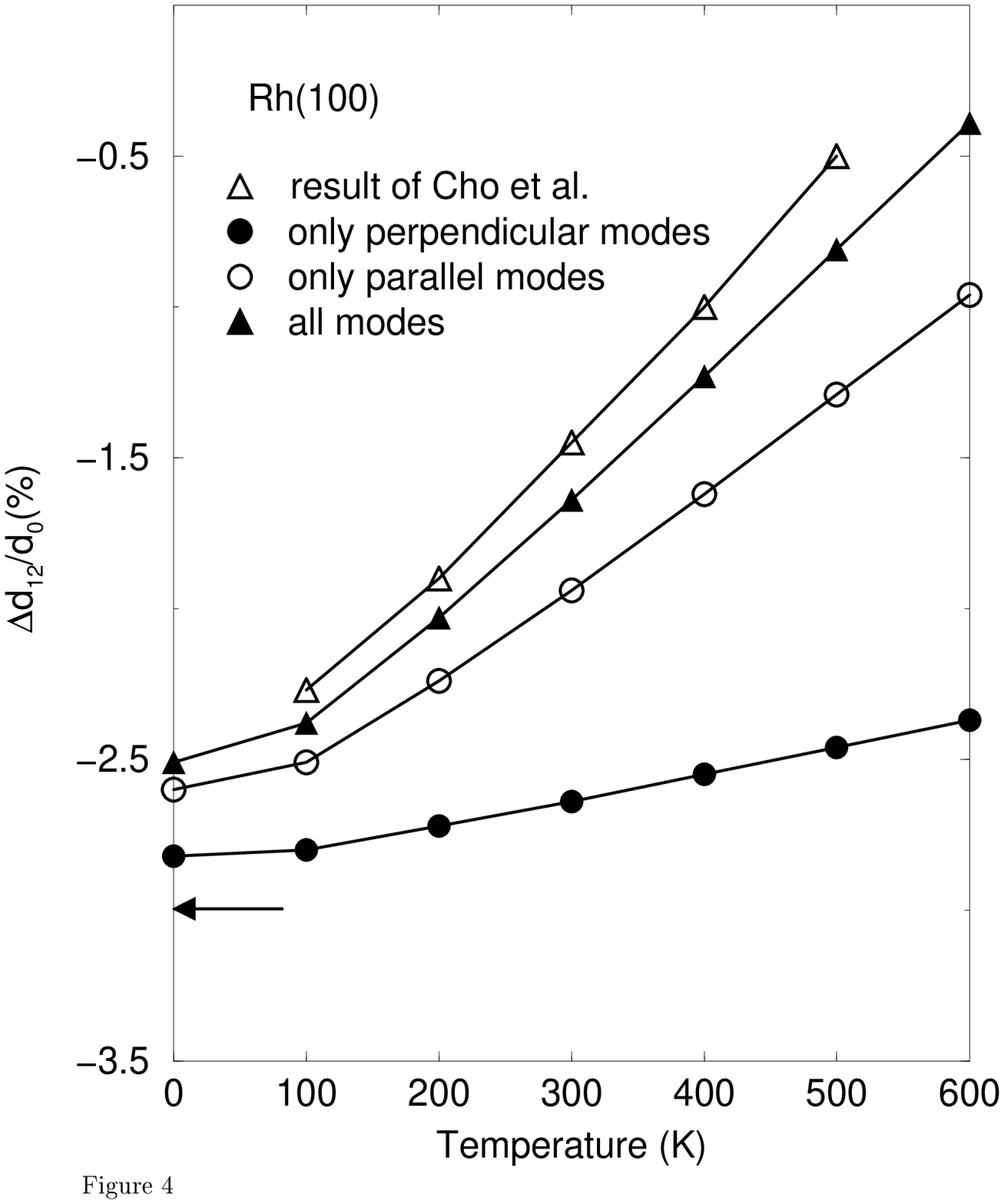,width=8cm}
\caption{ Top-layer relaxation of Rh(100) as a function of temperature
calculated by including the vibration modes at  $\overline{\Gamma}$,
 $\overline{\rm X}$, and $\overline{\rm M}$.
The contributions from  vibration parallel and perpendicular
to the surface are are displayed. The arrow points at the result 
obtained when
vibrational contributions to the free energy are neglected completely.
The results of Cho
and Scheffler~{\protect \cite{Cho}} are shown for comparison.}
\label{figure4}
\end{figure}

Using the ``frozen'' phonon 
method~\cite{Ho86,Cohen82,Srivastava}, it is not practicable to get the phonon
frequency for an arbitrary  ${\bf q}_\parallel$ point in SBZ.  
For phonon, in contrast to electrons,  it is in fact acceptable to approximate 
the summation over the whole SBZ by the contributions
from only the high-symmetry ${\bf q}_\parallel$ points. The weights
of  $\overline{\Gamma}$,
$\overline{\rm X}$, and $\overline{\rm M}$ are $\frac{1}{4}, \frac{2}{4}$,
and $\frac{1}{4}$, respectively.  Thus, the vibrational free energy 
is given by
$F_{\rm vib}=\frac{1}{4}F_{\rm vib}(\overline{\Gamma})
+\frac{2}{4}F_{\rm vib}(\overline{\rm X})
+\frac{1}{4}F_{\rm vib}(\overline{\rm M})$. 
The temperature dependence of the top-layer
relaxation for Rh(100) by including all the vibrational contributions
from $\overline{\rm X}, \overline{\rm M}$, and 
$\overline{\Gamma}$ is shown in Fig.~\ref{figure4} . 
The arrow marks the result of the surface relaxation obtained when lattice
vibrations (also zero-point effects) are neglected.
We find that at low temperature the thermal expansion 
for Rh(100) is close to the result of Ref.~\cite{Cho}.
At 300 K, the top-layer relaxation is $-1.64$\%, the
value of Ref.~\cite{Cho} is $-1.45$\%.  However,
for higher temperature the thermal expansion will be overestimated 
if only the  
vibrations  at $\overline{\Gamma}$ point are included. 
In general, our improved calculations
confirm the conclusion of Ref.~\cite{Cho}.  The calculated surface thermal
expansion coefficient is $\alpha_{s}=(d_{12})^{-1}(\partial d_{12}/\partial
T)=40.7\times 10^{-6}{\rm K}^{-1}$ at $T$=300 K, which is 5.0 times larger than
in the bulk ($\alpha_{b}=8.2\times 10^{-6}{\rm K}^{-1}$~\cite{CRC96}). 
Thus for Rh(100) the thermal expansion is clearly increased 
at the surface. 
The different contributions to the surface relaxation 
from vibrations parallel to the surface and perpendicular
to the surface are also shown in Fig.~\ref{figure4}.  
All the vibrational modes at $\overline{\rm X}$,
$\overline{\Gamma}$, and $\overline{\rm M}$ are included. 
The results in Fig.~\ref{figure4} confirm that the parallel vibrations 
have a significantly larger influence on the surface thermal expansion 
than the perpendicular vibrations ($z$-direction). This
is in agreement with the findings and conclusions of 
Refs.~\cite{Nara,Cho}.

\begin{figure}
\psfig{figure=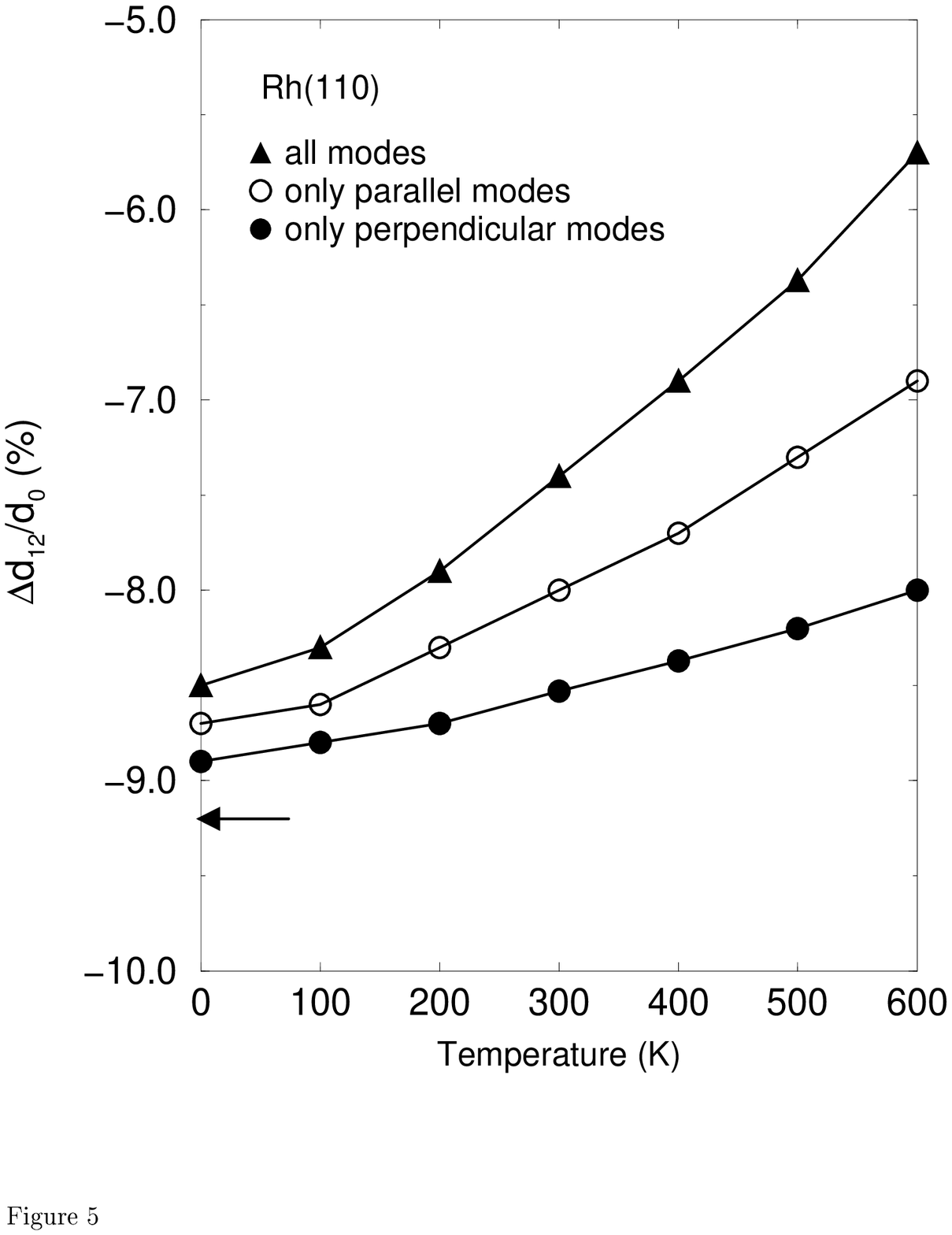,width=8cm}
\caption{ Top-layer relaxation of Rh(110) as a function of temperature
calculated by including the vibrational modes at $\overline{\Gamma}$,
$\overline{\rm X}$, $\overline{\rm Y}$, and $\overline{\rm S}$.
The contributions from vibrations parallel and perpendicular
to the surface are displayed. The arrow points at the result obtained 
when vibrational contributions to the free energy are neglected completely.
}
\label{figure5}
\end{figure}

We now turn to the more open (110) surface.
The temperature dependence of the top-layer relaxation for Rh(110) is
shown in Fig.~\ref{figure5}. The vibrational free energy is calculated
by summing over the wave vectors $\overline{\Gamma}$, $\overline{\rm X}$,
$\overline{\rm Y}$, and $\overline{\rm S}$ in the SBZ.
These four points all have the same weight, therefore the 
vibrational free energy is given by 
$F_{\rm vib}=\frac{1}{4}[F_{\rm vib}(\overline{\Gamma})+F_{\rm vib}
(\overline{\rm X})
+F_{\rm vib}(\overline{\rm Y})+F_{\rm vib}(\overline{\rm S})]$.
Similar to  Rh(100), the contribution to the thermal expansion of
$d_{12}$
from in-plane vibrations is larger than that from out-plane vibrations.
At $T$=300 K we obtain the surface thermal expansion coefficient $\alpha_{s}=59.4
\times 10^{-6} {\rm K}^{-1}$, which is 7.2 times larger than the 
bulk thermal expansion coefficient.
Comparing the thermal expansion coefficient of Rh(110) with that
of Rh(100), we find that Rh(110) exhibits more significant
anharmonicity than
Rh(100). The top-layer relaxation is $\Delta d_{12}/d_0 = -7.4$\%
at $T$=300 K which is close to the result of a
room temperature LEED analysis~\cite{Nichtl}, which gave a value
of $\Delta d_{12}/d_0 = -6.9$\%. The surface relaxation
obtained when just the total energy is minimized is
noticeably different, namely $\Delta d_{12}/d_0 = -9.2$\%.

\begin{figure}
\psfig{figure=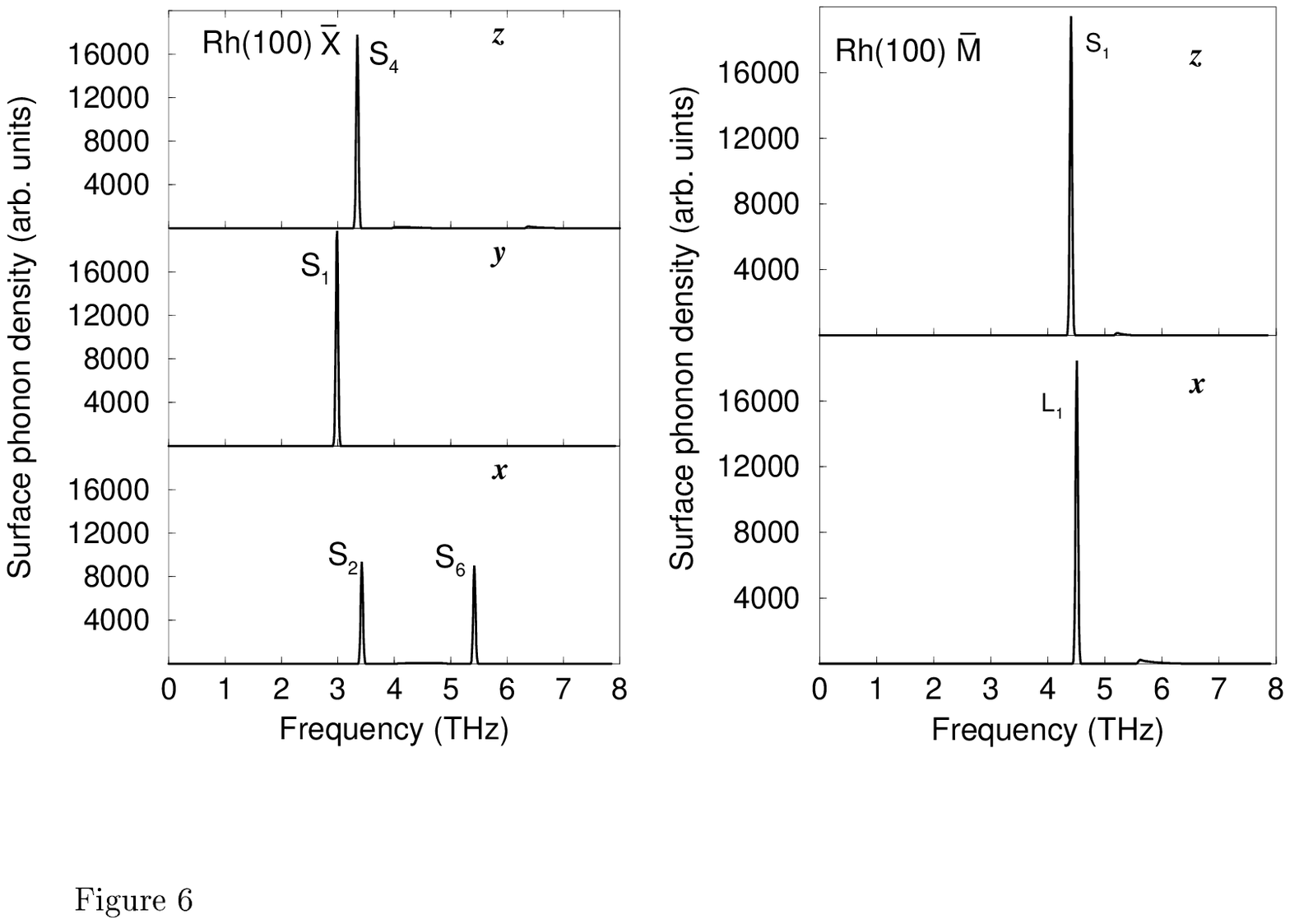,width=8cm}
\caption{ Surface phonon density of states of Rh(100) for the wave
vectors at the $\overline{\rm X}$
and $\overline{\rm M}$ points in the surface Brillouin zone. 
}
\label{figure6}
\end{figure}

\begin{figure}
\psfig{figure=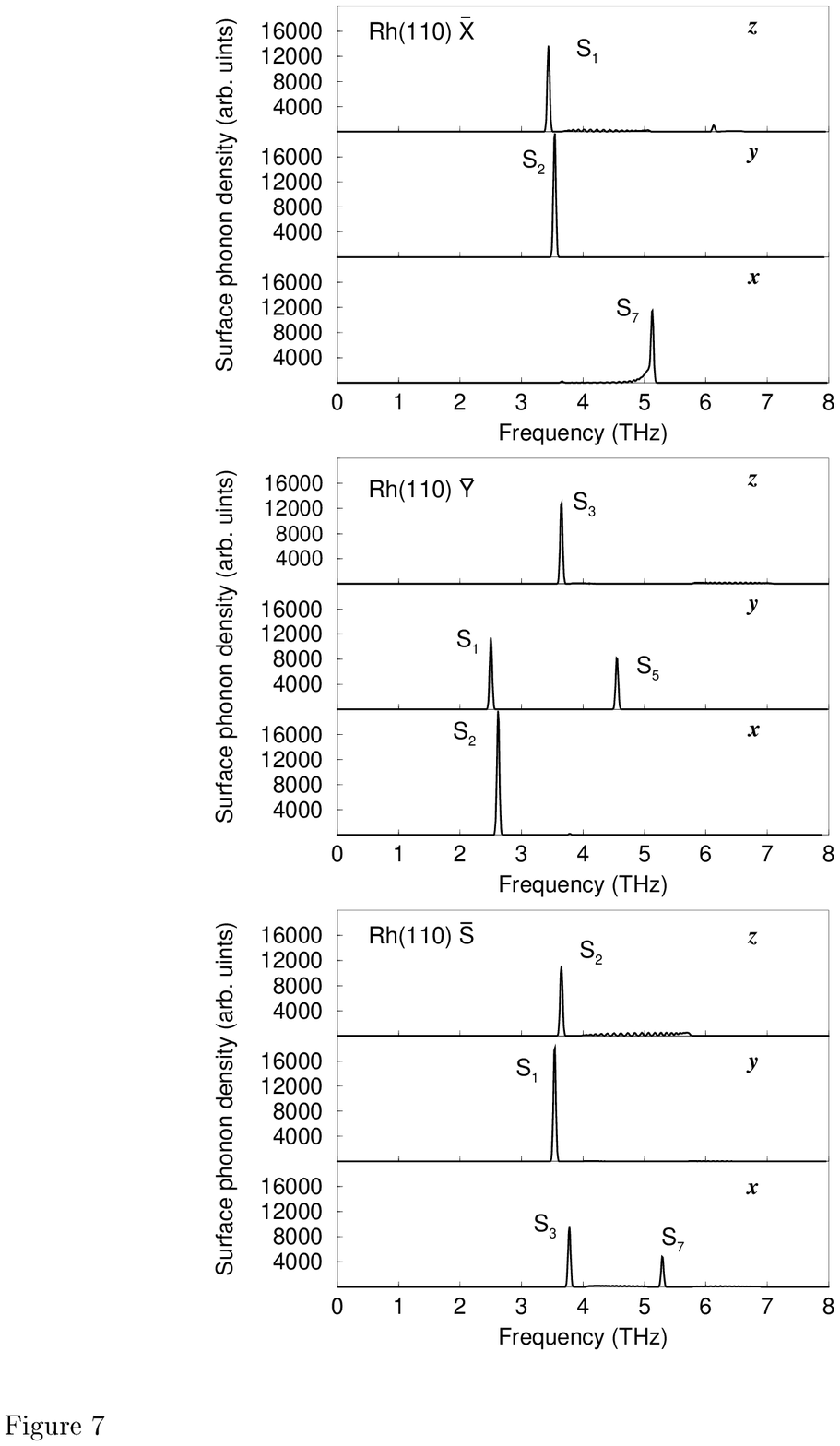,width=8cm}
\caption{ Surface phonon density of states of Rh(110) for the 
wave vectors at $\overline{\rm X}$
, $\overline{\rm Y}$, and $\overline{\rm S}$ points in the
surface Brillouin zone. }
\label{figure7}
\end{figure}

In Table \ref{table4}, we present the calculated surface phonon frequencies
corresponding to the surface relaxation at $T$=300 K. 
The  local density of states are obtained from
\begin{equation}
D_{{\bf q}_\parallel,\alpha}(\omega)=\sum_{p}\delta(\omega-\omega_{p})
|\zeta_{\alpha}(1, {\bf q}_\parallel, p)|^{2}
\end{equation}
where $\zeta_{\alpha}(1, {\bf q}_\parallel, p)$ is the surface layer polarization
vector for the $p$-th normal mode of the slab with respect to wave vector
${\bf q}_\parallel$, $\alpha$ denotes $x, y$ and $z$.
The calculated surface phonon densities of states are
shown in Fig.~\ref{figure6} and Fig.~\ref{figure7}. 
Different plots correspond to different polarizations.
At the $\overline{\rm M}$ point for Rh(100) 
( Fig.~\ref{figure6}), results for $y$ polarization are identical
to those for the $x$  polarization due to symmetry.
It can be seen that in all cases (Fig.~\ref{figure6} and Fig.~\ref{figure7}),
the surface modes are strongly localized at surface and are well 
separated from the bulk modes. 
Thus it should be possible to identify them in EELS 
or He scattering experiments. It is these localized surface phonons
that are most sensitive to the surface relaxations, and hence govern the 
anharmonic thermal expansion at the surface.
It can be seen in Figs.~\ref{figure6} and \ref{figure7}
that there are more localized surface 
phonon modes  polarized in the $x$ and $y$ directions than 
in the $z$ direction.
This is one of the reasons why the parallel vibrations give  a more
important contribution to the thermal expansion
 (see Figs.~\ref{figure4} and \ref{figure5}).  

\section{Summary}

In conclusion, we have used full-potential FP-LAPW
method to investigate the surface structures and surface phonon
spectra of Rh(100) and Rh(110) surfaces. We studied the thermal
expansion by taking anharmonicity into account
 within the quasiharmonic approximation.
It is found that different phonons give different contributions to 
the surface relaxation in the range of high temperature (higher than 300 K).
The surface thermal expansion coefficients calculated by including
all the high symmetry points in surface Brillouin zone
are 40.7$\times 10^{-6}{\rm K}^{-1}$ for Rh(100) and 
59.4$\times 10^{-6}{\rm K}^{-1}$
for Rh(110) at $T$=300 K, which are 5.0 and 7.2 times larger than
that of the bulk, respectively. The obtained surface relaxations of Rh(100)
and Rh(110) at $T=300$ K are in agreement with the experimental
measurements at room temperature.
The calculated results  confirm that the in-plane vibrations
give a more important contribution to the surface thermal expansion
than the out-of-plane vibrations. 
\acknowledgments

One of the authors (J.J. Xie) would like to acknowledge the financial support
from Alexander von Humboldt foundation in Germany.



\newpage

\begin{table}
\caption{Surface relaxations $\Delta d_{12}/d_{0}$ and  
$\Delta d_{23}/d_{0}$, work function $\phi $ (eV) and surface energies
$\sigma $ (eV/atom) for Rh(100) and Rh(110) surfaces
obtained by different calculations and experiments.
The results correspond to $T = 0$ K, and none of the 
theoretical values includes the effect of zero-point vibrations.
\label{table1}}
\begin{tabular}{l lcccc}
&  &$\Delta d_{12}/d_{0}$ &$\Delta d_{23}/d_{0}$ & $\phi $ & $\sigma $\\
\hline
Rh(100)  &LDA~\cite{Schef92}&  $-3.5$\% & --& 5.25 & 1.27   \\
 &LDA~\cite{Eichler96} &  $-3.8$\% & +0.7\%& --  & 1.44 \\
 &LDA~\cite{Cho} &  $-3.0$\% & $-0.2$\%& 5.26 & 1.29 \\
 &GGA~\cite{Cho} &  $-2.8$\% & $-0.1$\%& 4.92 & 1.04\\
 &this - LDA &  $-3.3$\% & $-0.2$\%& 5.30 & 1.23 \\
 &this - GGA &  $-3.0$\% & $-0.1$\%& 4.91 &  0.99 \\
 &experiment&  +0.5$\pm$1.0\%~\cite{Muller}&0$\pm$1.5\%~\cite{Muller}&
 4.65~\cite{American} & 1.12~\cite{Tyson}\\
 &experiment& $-1.16$$\pm$1.6\%~\cite{Begley}& 0$\pm$1.6\%~\cite{Begley}& \, & \, \\
\\
Rh(110)& LDA~\cite{Stokbro} &$-10.2$\% &+2.5\% & 4.99 & 1.92  \\
       & LDA~\cite{Eichler96} &$-9.8$\% &+2.6\% & --   & 2.05  \\
       & this - LDA &$-9.9$\% &+2.0\% &4.94& 1.85\\
       & this - GGA &$-9.2$\%  &+2.1\% &4.59 &1.43\\
 &experiment& $-6.9 \pm$1.0\%~\cite{Nichtl}& +1.9$\pm$1.0\%~\cite{Nichtl}& 
4.98~\cite{CRC} & 1.27~\cite{Mezey}\\
\end{tabular}
\end{table}

\begin{table}
\caption{ Interplanar force constants of Rh(100) coupling the surface 
layer to other layers for wave vectors at $\overline{\Gamma}$, 
$\overline{\rm X}$, and $\overline{\rm M}$ at 
$\Delta d_{12}$/$d_{0}=-3.0\%$, $\Delta d_{23}/d_{0}=-0.1\%$.
The units are $10^{4}$ dyn/cm. The corresponding interplanar
force constants for the interior layers of the slab are also shown
for comparison. 
The indices $(\alpha \beta)$ and the arguments $(l_3, l_3^\prime)$
of $\phi_{\alpha \beta}^{p}(l_{3},l_{3}^{\prime})$
are combined as $(l_{3}\alpha,l_{3}^{\prime}\beta)$. 
Matrix elements not listed are zero by symmetry.
\label{table2}}
\begin{tabular}{c  c c c c c }
$(l_{3}\alpha,l_{3}^{\prime}\beta)$  & Surface & Interior & 
$(l_{3}\alpha,l_{3}^{\prime}\beta)$  & Surface & Interior\\
\hline
        &         &    &${\bf q}_\parallel=\overline{\Gamma}$&    &    \\
$(1x,2x)$ & $-4.89$  &$-4.92$ &$(1x,3x)$& $-0.03$  &$-0.30$     \\
$(1z,2z)$ & $-8.34$  &$-7.65$ &$(1z,3z)$& $-1.45$  &$-1.84$    \\
\\
        &         &    &${\bf q}_\parallel=\overline{\rm X}$ &      & \\
$(1x,1x)$ &14.65 &26.23  &$(1x,3x)$&  0.53  &\,0.39 \\
$(1y,1y)$ & 6.34 &10.18  &$(1y,3y)$& $-0.08$  &\, 0.03 \\
$(1z,1z)$ &10.15 &17.07  &$(1z,3z)$& $-0.57$  &$-0.70$ \\
$(1x,2z)$& $-6.42$  &$-5.46$ &$(1z,2x)$& $-5.41$  &$-5.46$\\
\\
        &         &     & ${\bf q}_\parallel=\overline{\rm M}$&      &        \\
$(1x,1x)$& 14.60    & 23.70 & $(1x,3x)$ &$-0.50$&$-0.72$ \\
$(1z,1z)$& 15.01    & 18.49 & $(1z,3z)$ &$-0.88$   &$-0.59$\\
$(1x,2y)$& -2.59   &-2.73   & $(1y,2x)$ &$-2.59$ & $-2.73$ \\
\end{tabular}
\end{table}

\begin{table}
\caption{
Interplanar force constants for Rh(110) at $\overline{\Gamma}$,
$\overline{\rm X}$, $\overline{\rm Y}$, and $\overline{\rm S}$ at  
$\Delta d_{12}$/$d_{0} = -9.2$\%, $\Delta d_{23}/d_{0}= 2.1\%$.
The units are  $10^{4}$ dyn/cm. 
The corresponding interplanar
force constants for the interior layers of the slab are also shown
for comparison. 
The indices $(\alpha \beta)$ and the arguments $(l_3, l_3^\prime)$
of $\phi_{\alpha \beta}^{p}(l_{3},l_{3}^{\prime})$
are combined as $(l_{3}\alpha,l_{3}^{\prime}\beta)$.
Matrix elements not listed are zero by symmetry.
\label{table3}}
\begin{tabular}{c  c c c c c }
$(l_{3}\alpha,l_{3}^{\prime}\beta)$ & Surface & Interior
 & $(l_{3}\alpha,l_{3}^{\prime}\beta)$ & Surface & Interior\\
\hline
        &         &    &${\bf q}_\parallel=\overline{\Gamma}$&    &    \\
$(1x,2x)$ & $-5.13$  &$-4.30$ &$(1x,3x)$& 0.31   &$-0.86$     \\
$(1y,2y)$ & $-7.88$  &$-7.14$ &$(1y,3y)$& 0.15   &$-0.14$     \\
$(1z,2z)$ & $-3.95$  &$-3.55$ &$(1z,3z)$& $-6.87$  &$-4.44$    \\
\\
        &         &    &${\bf q}_\parallel=\overline{\rm X}$ &      & \\
$(1x,1x)$ & 13.52   &25.40 &$(1x,3x)$&0.71&0.62\\
$(1y,1y)$ &  9.51   &17.24 &$(1y,3y)$&$-0.07$&$-0.96$\\
$(1z,1z)$ & 13.68   &16.56 &$(1z,3z)$&$-5.88$&$-3.61$\\
$(1x,2z)$ & $-1.77$   &$-3.09$ &$(1z,2x)$&$-3.48$&$-3.09$\\
\\
        &         &    &${\bf q}_\parallel=\overline{\rm Y} $ &      & \\
$(1x,1x)$ & 5.69   &10.81 &$(1x,3x)$&$-0.49$& 0.11\\
$(1y,1y)$ & 9.11   &20.50 &$(1y,3y)$&1.74 & 1.08\\
$(1z,1z)$ &13.68   &18.12 &$(1z,3z)$&$-5.78$&$-3.67$\\
$(1y,2z)$ &$ -5.52$  &$-5.81$ &$(1z,2y)$&$-5.52$&$-5.81$\\
\\
        &         &     & ${\bf q}_\parallel=\overline{\rm S}$&      &        \\
$(1x,1x)$&13.83    &24.16 & $(1x,3x)$& 1.74& 0.94  \\
$(1y,1y)$&10.21    &\,17.81 &$ (1y,3y)$&0.76 & 0.25  \\
$(1z,1z)$&13.66    &\,16.44 & $(1z,3z)$&$-6.20$&$-2.92$  \\
$(1x,2y)$&$-6.12$   &$-5.23$  &$(1y,2x)$& $-4.23$&$-5.23$  \\
\end{tabular}
\end{table}

\vspace{2.5cm}
\begin{table}
\caption{
Calculated surface phonon frequencies $\nu$ for Rh(100) and Rh(110)
corresponding to surface relaxation at $T$=300 K. Units are in THz.
}
\label{table4}
\begin{tabular}{c c c c| c c c c c c}
 \multicolumn{4}{c|}{Rh(100)}& \multicolumn{6}{c}{Rh(110)}\\
\multicolumn{2}{c}{$\overline{\rm X}$} & \multicolumn{2}{c|}{$\overline{\rm 
M}$}& \multicolumn{2}{c}{$\overline{\rm X}$}&
\multicolumn{2}{c}{$\overline{\rm Y}$}&\multicolumn{2}{c}{$\overline{\rm S}$}\\
$\nu$& mode&$\nu$&mode&$\nu$&mode& $\nu$& mode& $\nu$& mode\\
\hline
2.97 &S$_{1}$& 4.44 &S$_{1}$ & 3.44 &S$_{1}$& 2.50 &S$_{1}$& 3.54 &S$_{1}$\\
3.36 &S$_{4}$& 4.50 &L$_{1}$ & 3.54 &S$_{2}$& 2.62 &S$_{2}$& 3.66 &S$_{2}$\\
3.42 &S$_{2}$&      &        & 5.13 &S$_{7}$& 3.65 &S$_{3}$& 3.78 &S$_{3}$\\
5.40 &S$_{6}$&      &        &      &       & 4.52 &S$_{5}$& 5.26 &S$_{7}$\\
\end{tabular}
\end{table}

\end{document}